\documentclass[ aip,amsmath,amssymb, reprint,]{revtex4-1}
\usepackage{graphicx}
\usepackage{dcolumn}
\usepackage{bm}
\usepackage[utf8]{inputenc}
\usepackage[T1]{fontenc}
\usepackage{mathptmx}
\usepackage{etoolbox}
\usepackage{color}
\makeatletter
\def\@email#1#2{%
 \endgroup
 \patchcmd{\titleblock@produce}
  {\frontmatter@RRAPformat}
  {\frontmatter@RRAPformat{\produce@RRAP{*#1\href{mailto:#2}{#2}}}\frontmatter@RRAPformat}
  {}{}
}%
\makeatother
\begin{document}

\preprint{AIP/123-QED}

\title{Entrainment of magnetic fluid by a moving boundary of a plane gap}
\author{D. S. Goldobin}
\email{denis.goldobin@gmail.com}
\author{Yu. L. Raikher}
\homepage{https://www.icmm.ru/~raikher.html}
\affiliation{Institute of Continuous Media Mechanics, Russian Academy of Sciences, Ural Branch, Perm, 614018, Russia}

\date{\today}

\begin{abstract}
     A fluid mechanics problem is solved which technological prototype is a fluid acoustic contact that is an inherent element of ultrasonic non-destructive testing procedures.
     It is well known that the acoustic contact established with an ordinary fluid suffers from essential disadvantage that is the loss of stability due to the gravity-induced fluid leakage in the course of dynamic scanning.
     The use of magnetic fluid (MF) is one of the ways to resolve the issue.
     A compact portion of MF held in place by a permanent magnet enables one to maintain a stable acoustic contact (fluid bridge) under arbitrary orientation of the ultrasonic sensor and, simultaneously, to radically minimize the drain of the contact fluid.
     The model system under consideration comprises a MF bridge that fills a flat gap, one of whose boundaries moves with constant velocity.
     Due to its wetting by the fluid, the receding plane carries away a fluid film thus depleting the contact.
     Theoretical expressions are obtained which define the profile of the film in the dynamic regime and the dependencies of the magnetic fluid drain on the boundary velocity, gap height and configuration of the imposed magnetic field.
     On that basis the optimal parameters are evaluated which ensure effective retention of the fluid contact under minimal drain of the fluid from it.
\end{abstract}

\maketitle

\section{Introduction \label{sec:1}}
     The generic principle of ultrasonic non-destructive testing, from industry to medicine, is echo-probing.
     A piezoelement excited by an electric pulse emits an ultrasonic wave package and sends it off into the tested sample.
     Then the piezoelement switches into the receiving mode, gets the acoustic signal (echo) reflected by the defect(s) and converts it into an electric signal.
     This latter is analyzed and processed yielding information on the presence and location of defects.

     A crucial parameter for the quality of control is the energy of the probing signal pumped in the item.
     A standard way to ensure effective transfer, is to fill the gap between the sensor and the surface of the tested item by an appropriate fluid, i.e., establish a fluid bridge.
     This brings the values of wave resistances of the acoustic path elements closer and, at the same time, does not impede motion of the sensor along the tested surface (dynamic scanning).

     To establish acoustic path, the immersion fluid must be wetting both boundaries of the contact gap.
     This entails that, while scanning, the controlling unit leaves behind it a fluid film of a finite thickness thus causing the contact fluid loss.
     In the case of ordinary fluid (water, alcohol, glycerol) there is, however, a fat chance that the contact might fail due to massive drain caused by the gravity force.
     That is why, the problem of maintaining integrity of the fluid bridge, i.e., stable acoustic contact, is of high importance for the ultrasonic control technology.
     In the present work we consider an unconventional solution of the problem that grants stability of the contact at arbitrary direction of the gravity force and ensures minimal loss of the contact fluid.

\section{Magnetofluidic acoustic contact \label{sec:2}}
     The key idea is to establish acoustic contact with the aid of magnetic fluid (MF).
     A magnetic mounting juxtaposed with the sensor creates around the latter a gradient field that attracts MF and keeps it in place.
     Provided the magnet is sufficiently strong, this may effectively prevent the gravity-induced break (leak) of the acoustic path.

     This concept had been put forward for the first time in 1990's as one of practical applications of MF \cite{KoSi_ITMO_83,PrSi_Pat_83} and for some time had been developing just empirically \cite{BrKo_Pat_84,PrBa_SJNDT_85,AlBa_SJND_85,KoRa_Pat-85,KoRa_Pat_87,KoKo_SJNDT_88,KoLe_SJNDT_89,KoRa_E_92,KoRa_EJRNDT_92}.
     Meanwhile, it is evident that its successful implementation to engineering practice requires that a number of problems of fundamental fluid mechanics of MF films were solved.
     One of those is addressed in the present paper.

\begin{figure}[ht]
\includegraphics[angle=90,width=0.4\textwidth]{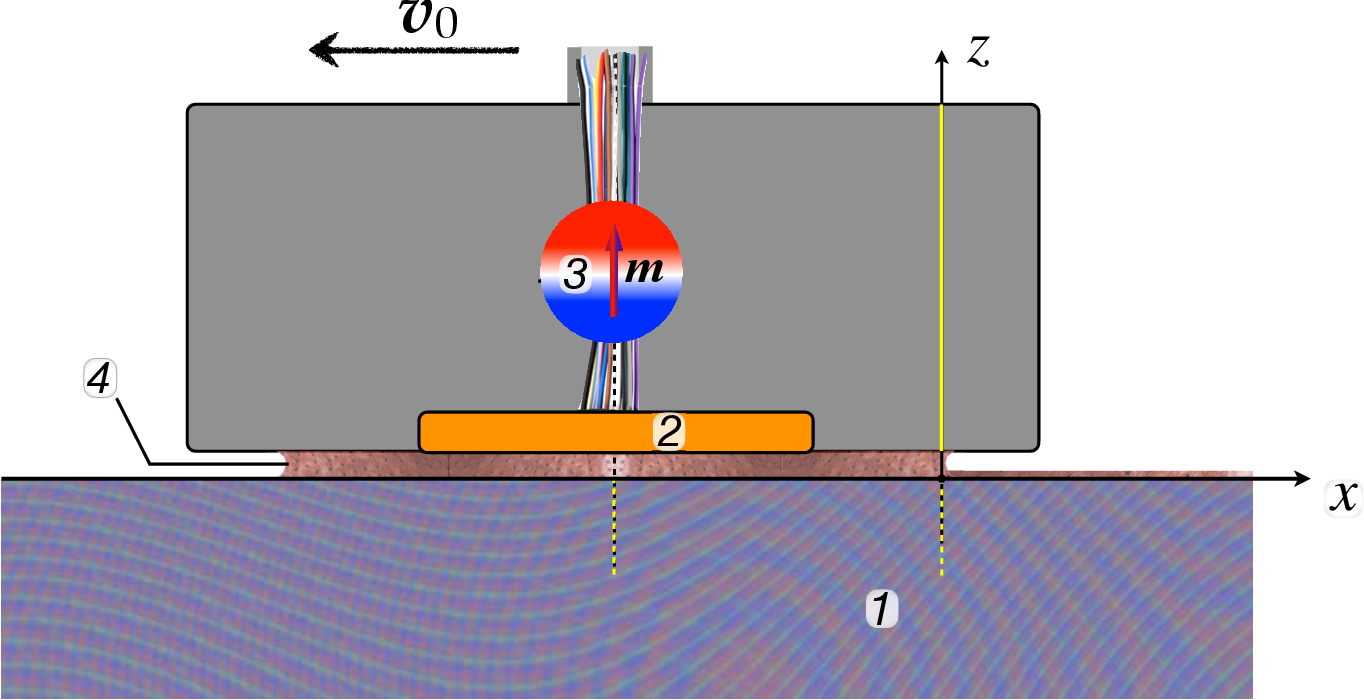}
\caption{Schematic illustration of the ultrasonic scanning process with an MF acoustic contact: (\emph{1})~tested item, (\emph{2})~ultrasonic sensor, (\emph{3})~magnetic mounting, (\emph{4})~magnetic fluid (fluid bridge).
\label{fig:01}}
\end{figure}

     A scheme of an ultrasonic scanning unit with a magnetofluidic acoustic contact is shown in Figure \ref{fig:01}.
     A sensor that contains a piezoelement and is equipped with a permanent magnet is positioned at a certain height above the surface of the tested item.
     The MF is located inside the in-between gap and forms a fluid bridge that fully covers the emitting face of the sensor.
     The sensor and item are set to relative motion with velocity $\bm{v}_0$.

     When performing scanning, the contact wets the surface of the item and leaves on it a residual MF film.
     Close to the rear edge of the fluid bridge, where the film yet forms, the main factors affecting its profile are the magnetic force and surface tension.
     Meanwhile, far from the contact the gradient magnetic field becomes negligible, and the main factors affecting the film profile are the surface tension and viscosity.

     Note that the dynamics of MFs under conditions like the ones presented in Figure \ref{fig:01} up to now had hardly been touched.
     As far as we know, the MF film science had been developing along two main lines, namely: (1)~the effect of applied uniform field on the free flow of MF down an inclined plane, see, e.g., \cite{ReSu_PF_94,CoMa_PF_15,CoMa_JEM_17,YaLi_PF_20}, and (2)~a forced flow of MF in thin closed gaps where it works as a lubricant, see, e.g., \cite{OsNa_TL_01,ChHu_TI_14,StSe_CP_24}.
     The problem in question, being quite far from the above-mentioned topics, seems very much alike those of the \emph{slot-die-coating} type, see \cite{MaCa_AIChE_00,LFC_18,DiLi_AIChE_16}, for example.
     However, to the best of our knowledge, in the accessible literature there is not a single work where the coating fluid had been considered as magnetic.

     The basics of the \emph{slot-die-coating} approach had been put forward in the classic paper \cite{LaLe_ActaUSSR_42} (reprinted in \cite{,LaLe_DCF_88}).
     Its theoretical formulation was inspired by the necessity to understand and describe the process of coating a polymeric film with a photoemulsion layer.
     On the one hand, the present problem is akin to this situation since here we also consider the process where a fluid layer with a free surface is forcibly carried away by a moving plane.
     On the other hand, the differences are evident as well.
     Whereas in the case of \cite{LaLe_ActaUSSR_42} gravitation makes the fluid to flow down, in the ultrasonic scanning of a horizontal surface the gravity, at the most, helps to prevent rippling of the MF surface.
     The essential difference, however, is in the profile of the fluid at the rear edge of the fluid bridge (dynamic contact angle).
     In the classic situation, the meniscus is formed due to interplay of gravity and capillary forces and nothing else.
     In the here considered case, the main effect on the fluid free surface in the `near' zone is due to the magnetic (ponderomotive) forces, and, because of that, as shown below, the magnetic effect on the flow may easily become decisive.

\section{Residual magnetic fluid film \label{sec:3}}
\begin{figure}[ht]
\includegraphics[angle=90,width=0.3\textwidth]{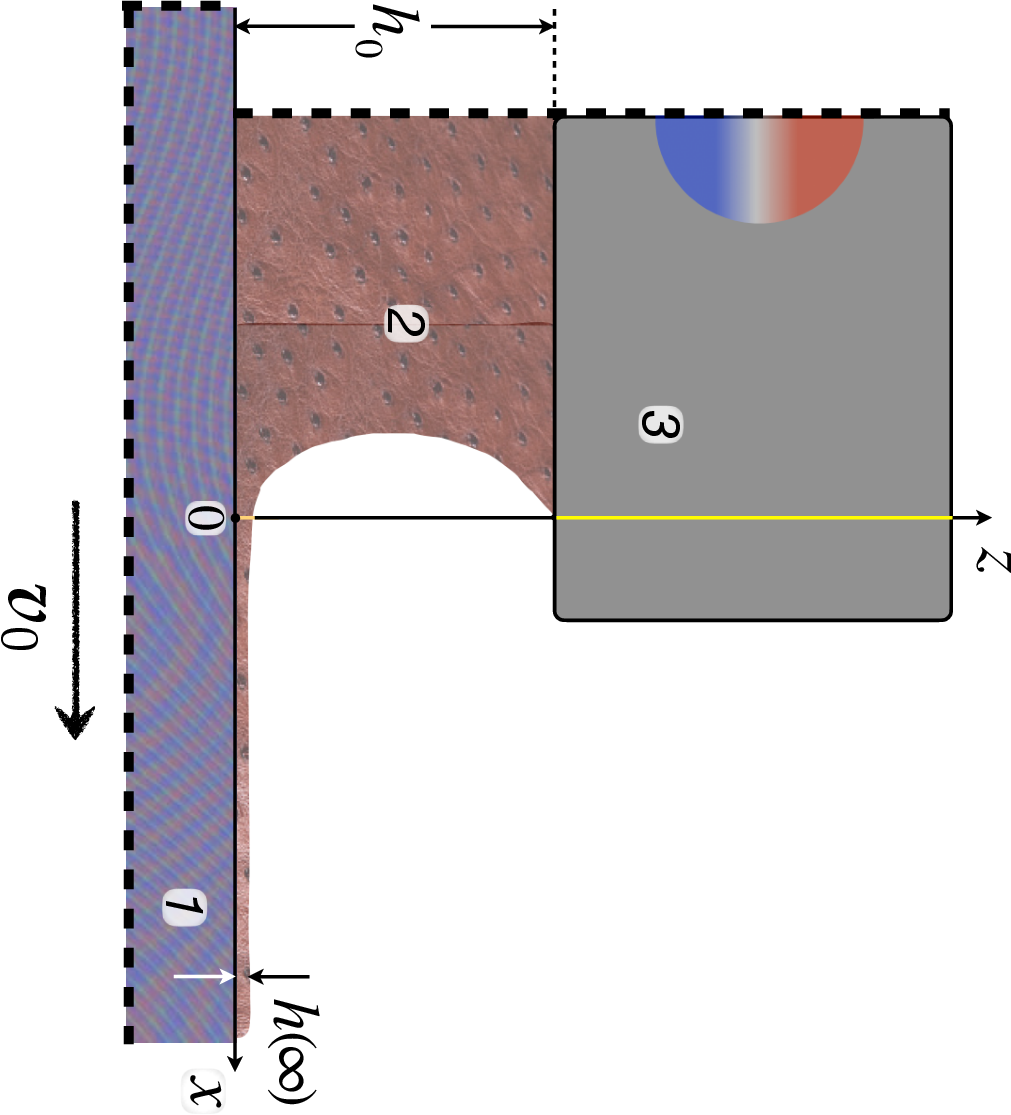}
\caption{On the 2D problem of entrainment of MF by a moving wetted boundary of a plane gap; (\emph{1}) lower boundary of the gap (moving), (\emph{2})~magnetic fluid (fluid bridge), (\emph{3}) upper boundary of the gap (fixed). Black dashed lines indicate that only a part of the whole scheme is shown.
\label{fig:02}}
\end{figure}

     A simplified sketch of the system under consideration is shown in Figure \ref{fig:02}.
     The problem is formalized as follows.
     The geometry is 2D, where the $Oz$ axis is directed normally to the scanned surface, and the $Ox$ axis is parallel to it.
     The gap thickness $h_0$ is assumed to be much greater than $h(\infty)$, that is the film thickness far from the gap.
     In this approximation, the fluid surface profile near the rear edge of the fluid bridge ($x=0$) might be treated as being very close to the static meniscus.

     In the coordinate frame fixed to the sensor, the scanned surface goes away with constant velocity $\bm{v}_0$.
     Just beyond the rear edge of the fluid bridge, the meniscus profile quite sharply transforms into a layer-like (film) profile making the dynamic contact angle, that is, the thickness $z=h(x)$ of the receding film rapidly decreases with the distance $x$; at $x\gg h_0$ the thickness $h(x)$ asymptotically tends to a constant value $h(\infty)$.

     Following the Landau-Levich approximation \cite{LaLe_ActaUSSR_42}, under assumption that the fluid velocity is not high, the fluid profile is divided in two zones: the `far' one where the fluid surface is virtually horizontal, and the `near' one where the profile of dynamic meniscus is close to that of the static one.

     Keeping in mind an MF of the magnetite-based type \cite{Rosen_FHD_85}, we treat it as an isotropic medium, magnetizable with the susceptibility $\chi$, i.e., assume a linear magnetization law:
\begin{equation} \label{eq:01}
    M_i=\chi H_i.	
\end{equation}

     In the `near' zone where the field is strong, one should add the Maxwell stress tensor to the usual thermodynamic pressure $p$ in MF.
     The latter is taken in the form corresponding to the absence of electric field and current \cite{Rosen_FHD_85}.
     Accordingly, the force volume density in the fluid may be presented as
\begin{equation*}
   \nabla_k\sigma_{ik}=-\nabla_ip+\mu_0\nabla_i\int_0^H M(H^\prime)dH^\prime,
\end{equation*}
that is valid for any unhysteretic magnetization law.

     The components of magnetic field at the interface between MF (that obeys the law (\ref{eq:01})) and the non-magnetic medium ($\chi=0$) are related as
\begin{equation}  \label{eq:02}
   H_\tau=H_\tau^{\rm (e)}, \qquad H_n=\frac{1}{1+\chi}H_n^{\rm (e)};
\end{equation}
where $\mu_0$ is vacuum magnetic permeability.
     Here subscripts $\tau$ and $n$ denote, respectively, the directions tangential and normal to the interface whereas superscript ($\rm e$) marks the external non-magnetic medium (air).
     In equilibrium, the tangential stresses are absent whereas when passing from the non-magnetic medium into MF there occurs a pressure jump entailed by the joint action of surface tension and magnetic field:
\begin{equation}  \label{eq:03}
   \Delta p=p_\sigma-\tfrac12\mu_0M_n^2=p_\sigma-\tfrac12\mu_0\chi^2H_n^2;
\end{equation}
here $p_\sigma$ is capillary pressure and $M_n$ and $H_n$ are the normal components of magnetization and internal field; in the last part of equation (\ref{eq:03}) formula (\ref{eq:01}) has been used.
     In what follows the so-called induction-free approximations is used assuming that $\chi\ll1$; this enables one not to take into account the demagnetising fields.

\subsection{Fluid flow in the `far' zone  \label{sec:3.1}}
     In the `far' zone the major velocity component is $v_x$ that enables one to substantially simplify the pertinent fluid dynamics equations.
     As the velocity gradient $\partial v_x\big/\partial z$ is much greater than $\partial v_x\big/\partial x$, to describe the motion of the fluid that resides on a flat surface it suffices to take the Prandtl boundary-layer equations.
     For a steady flow of MF they take the form
\begin{equation}
\label{eq:04}
\begin{array}{l}
\displaystyle
   \eta\frac{\partial^2v}{\partial z^2}\!=\!\frac{\partial p}{\partial x}\!-\!\mu_0\frac{\partial}{\partial x}\int\limits_0^H M(H^\prime)dH^\prime\!\approx\!\frac{\partial p}{\partial x}\!-\!\tfrac12\mu_0\chi\frac{\partial H^2}{\partial x}, \\
\displaystyle
   \frac{\partial p}{\partial z}\!=-\rho g+\!\mu_0\frac{\partial}{\partial z}\int\limits_0^H M(H^\prime)dH^\prime\!\approx\!-\rho g+\tfrac12\mu_0\chi\frac{\partial H^2}{\partial z},
\end{array}
\end{equation}
where
\begin{equation*}
   H^2=\left(H_x^{\rm (e)}\right)^2+\left(\frac{H_z^{\rm (e)}}{1+\chi}\right)^2;
\end{equation*}
here $\eta$ is the dynamic viscosity coefficient and $g$ the gravity acceleration; from now on $v_z$, being the only non-negligible velocity component, is denoted as $v$.
     We remark that unlike the `far'-zone flow considered in \cite{LaLe_ActaUSSR_42}, in (\ref{eq:04}) gravitation enters the equation for $z$- and not for $x$-component.

     The boundary conditions for equations (\ref{eq:04}) are as follows.
     At the moving surface, i.e., for all $x$'s, there is no slippage of the fluid:
\begin{equation}  \label{eq:05}
v=v_0 \quad {\rm for} \quad z=0.
\end{equation}
     At the free surface, $z=h(x)$, the pressure inside MF equals the sum of atmospheric one (that might be taken as the reference level) and the pressure jump (\ref{eq:03}), the tangential stresses are absent:
\begin{equation}  \label{eq:06}
   p=p_\sigma -\mu_0\chi^2\frac{\left[H_z^{\rm(e)}(z=h)\right]^2}{2(1+\chi)^2}, \quad \eta\frac{\partial v}{\partial x}=0 \quad {\rm for} \quad z=h(x);
\end{equation}
note that the second formula is valid for a thin film ($\partial h\big/\partial x\ll1$).

     In the 2D problem formulation the capillary pressure is
\begin{equation}  \label{eq:07}
   p_\sigma=-\sigma\big/R,
\end{equation}
where $\sigma$ is surface tension coefficient and $R$ curvature radius of the meniscus in the $xOz$ plane.
     Substituting the standard expression for the curvature radius into (\ref{eq:07}) yields
\begin{equation}  \label{eq:08}
   p_\sigma=-\sigma\frac{d^2h}{dx^2}\left[1+\left(\frac{dh}{dx}\right)^2\right]^{-3/2}.
\end{equation}
For a nearly flat film the square of the first derivative in (\ref{eq:08}) may be neglected, so that the expression reduces to
\begin{equation}  \label{eq:09}
p_\sigma=-\sigma\frac{d^2h}{dx^2}.
\end{equation}
     With allowance for equations (\ref{eq:04}) and (\ref{eq:09}) the first one of the boundary conditions (\ref{eq:06}) takes the form
\begin{multline}  \label{eq:10}
   p=-\sigma\frac{d^2h}{dx^2}-\mu_0\chi^2\frac{\left[H_z^{\rm(e)}(z=h)\right]^2}{2(1+\chi)^2}-\rho g\left[z-h(x)\right] \\
+\tfrac12\mu_0\chi\left[H^2(x,z)-H^2(x,z=h(x))\right] \\
\approx -\sigma\frac{d^2h}{dx^2}\!-\!\rho g\left[z-h(x)\right] \\
+\!\frac{\mu_0\chi}{2}\!\!\!\left[\!\frac{\partial H^2}{\partial z}\!\left[z\!-\!h(x)\right]\!-\!\frac{\chi\left[H_z^{\rm(e)}(z=h)\right]^{\!2}}{(1+\chi)^2}\!\right].
\end{multline}
Here and below we make use of the smallness of derivatives $\partial H^2\!\!\big/\partial x$ and $\partial H^2\!\!\big/\partial z$: $h|\nabla H^2|\ll H^2$.

     As it is seen from (\ref{eq:10}), derivative $\partial p\!\!\big/\partial x$ and the other contributions in the right-hand part of the first of equations (\ref{eq:04}), to the leading order, do not depend on the $z$-coordinate:
\begin{multline*}
   \frac{\partial p}{\partial x}=\rho g\frac{dh}{dx}-\sigma\frac{d^3h}{dx^3} \\
-\tfrac12\mu_0\chi\!\!\left[\frac{dh}{dx}\frac{\partial H^2}{\partial z}\!+\!\frac{\chi}{(1\!+\!\chi)^2}\!\left(\!\frac{\partial}{\partial x}\!+\!\frac{dh}{dx}\frac{\partial}{\partial z}\!\right)\!\left(\!H_z^{\rm(e)}\!\right)^{\!\!2}\!\big|_{z=h(x)}\!\right] \\
\approx \rho g\frac{dh}{dx}-\sigma\frac{d^3h}{dx^3} \\
-\tfrac12\mu_0\chi\!\!\left[\frac{dh}{dx}\frac{\partial H^2}{\partial z}\!+\!\frac{\chi}{(1\!+\!\chi)^2}\!\!\frac{\partial(\!H_z^{\rm(e)}\!)^{2}}{\partial x}\!\Bigg|_{z=h(x)}\!\right].
\end{multline*}
     Two contributions (in square brackets) in the last expression are retained since, although $dh\big/dx$ and $\chi$ are small quantities, the hierarchy of their smallness is not yet established.
     Integration of the first boundary-layer equation of (\ref{eq:04}) after substituting boundary conditions (\ref{eq:05}), (\ref{eq:06}) and expression (\ref{eq:10}) for the pressure, yields the MF velocity in the form
\begin{multline}  \label{eq:11}
   v=v_0+\frac{1}{\eta}\left(\frac{\partial p}{\partial x}-\tfrac12\mu_0\chi\frac{\partial H^2}{\partial x}\right)\left(\tfrac12z^2-hz\right) \\
=v_0+\frac{1}{\eta}\left(\rho g_z\frac{dh}{dx}-\sigma\frac{d^3h}{dx^3} \right. \\
\left. -\tfrac12\mu_0\chi\frac{\partial}{\partial x}\left[(\!H_x^{\rm(e)}\!)^{2}+\frac{(\!H_z^{\rm(e)}\!)^2}{1+\chi}\right]\right)\left(\tfrac12z^2-hz \right),
\end{multline}
where notation $g_z=g-\tfrac12(\mu_0\chi\big/\rho)\left(\partial H^2\big/\partial z \right)$ is introduced, see also formulas (\ref{eq:23}) and (\ref{eq:24}) below.

     Due to the continuity equation, the thickness of the film retained by the moving surface defines the MF outflow (drain).
     Under steady conditions, at any distance $x$ the flux of the fluid per its unit width along $Oy$ is
\begin{equation}  \label{eq:12}
   j=\int_0^hv\,dz={\rm const}.
\end{equation}
     Substitution of $v$ from formula (\ref{eq:11}) enables one to express the flow in terms of the film thickness $h(x)$ in the `far' zone:
\begin{multline}  \label{eq:13}
   j=v_0h-\left(\rho g_z\frac{dh}{dx}-\sigma\frac{d^3h}{dx^3} \right. \\
\left. -\tfrac12\mu_0\chi\frac{\partial}{\partial x}\left[(\!H_x^{\rm(e)}\!)^{2}+\frac{(\!H_z^{\rm(e)}\!)^2}{1+\chi}\right]\right)\frac{h^3}{3\eta}.
\end{multline}
     Resolving equation (\ref{eq:13}) with respect to the higher-order derivative, one gets
\begin{equation}  \label{eq:14}
   \frac{d^3h}{dx^3}=\frac{3\eta}{\sigma}\frac{(j-v_0h)}{h^3}+\frac{\rho g_z}{\sigma}\frac{dh}{dx}-\frac{\mu_0\chi}{2\sigma}Q,
\end{equation}
where
\begin{equation*}
   Q\equiv\frac{\partial}{\partial x}\left[(\!H_x^{\rm(e)}\!)^{2}+\frac{(\!H_z^{\rm(e)}\!)^2}{1+\chi}\right].
\end{equation*}
     Passing then to dimensionless units
\begin{equation}  \label{eq:15}
   \lambda=\left(\frac{3\eta}{\sigma}\right)^{\!\!1/3}\frac{v_0^{4/3}}{j}x, \qquad \mu(\lambda)=\frac{v_0h(x)}{j}
\end{equation}
casts equation (\ref{eq:14}) into the form
\begin{equation}  \label{eq:16}
   \frac{d^3\mu}{d\lambda^3}-\frac{1-\mu}{\mu^3}-\frac{\rho g_zj^2}{(3\eta)^{2/3}\sigma^{\!1/3}v_0^{\!8/3}}\frac{d\mu}{d\lambda}+\frac{\mu_0\chi j^2}{6\eta v_0^3}Q=0.
\end{equation}
     Factor $Q$ in the last equation might be treated as a constant; to evaluate it, it suffices to take the field derivatives in the vicinity of the meniscus.

     The relevance of the last two terms in equation (\ref{eq:16})---they comprise all the dimensional parameters of the problem---is defined by the type of dependence $j(v_0)$.
     If the flux is proportional to $v_0$ in the power higher than 3/2, then the last term in (\ref{eq:16}) at sufficiently low velocities would be much below unity.
     Note that this condition coincides with that adopted in \cite{LaLe_ActaUSSR_42}.
     Provided it is satisfied, the next to last term in (\ref{eq:16}) would be small the more so because neglecting it requires a more mild condition: $j\propto v_0^{4/3+\varepsilon}$ is sufficient.

     Assuming that the flux indeed depends on the velocity as $j\propto v_0^\beta$ for $\beta>3/2$, one concludes that, provided the velocities is not too high, the last two terms in (\ref{eq:16}) may be neglected in comparison with the first two ones.
     In Section \ref{sec:6} below it will be demonstrated that this assumption holds in quite a wide interval of $v_0$ values.
     As a result, equation (\ref{eq:16}) reduces to
\begin{equation}  \label{eq:17}
   \frac{d^3\mu}{d\lambda^3}-\frac{1-\mu}{\mu^3}=0.
\end{equation}
     Note that equation (\ref{eq:17}) does not comprise any parameters that implies that for a given set of boundary conditions it should be solved just once.

     The boundary conditions for (\ref{eq:17}) are as follows.
     Far from the fluid bridge, i.e., at large $x$'s, the film thickness tends to a constant limit:
\begin{equation}  \label{eq:18}
   h(+\infty)=\lim_{x\rightarrow+\infty} h(x)=j/v_0,
\end{equation}
whereas the derivatives $dh/dx$ and $d^2h/dx^2$ tend to zero.
     In terms of the above-introduced dimensionless units the pertinent relationships assume the form
\begin{equation}  \label{eq:19}
\mu\rightarrow1, \quad \frac{d\mu}{d\lambda}\rightarrow0, \quad \frac{d^2\mu}{d\lambda^2}\rightarrow0 \quad {\rm for} \;\; \lambda\rightarrow+\infty.
\end{equation}

\begin{figure}[ht]
\includegraphics[angle=90,width=0.38\textwidth]{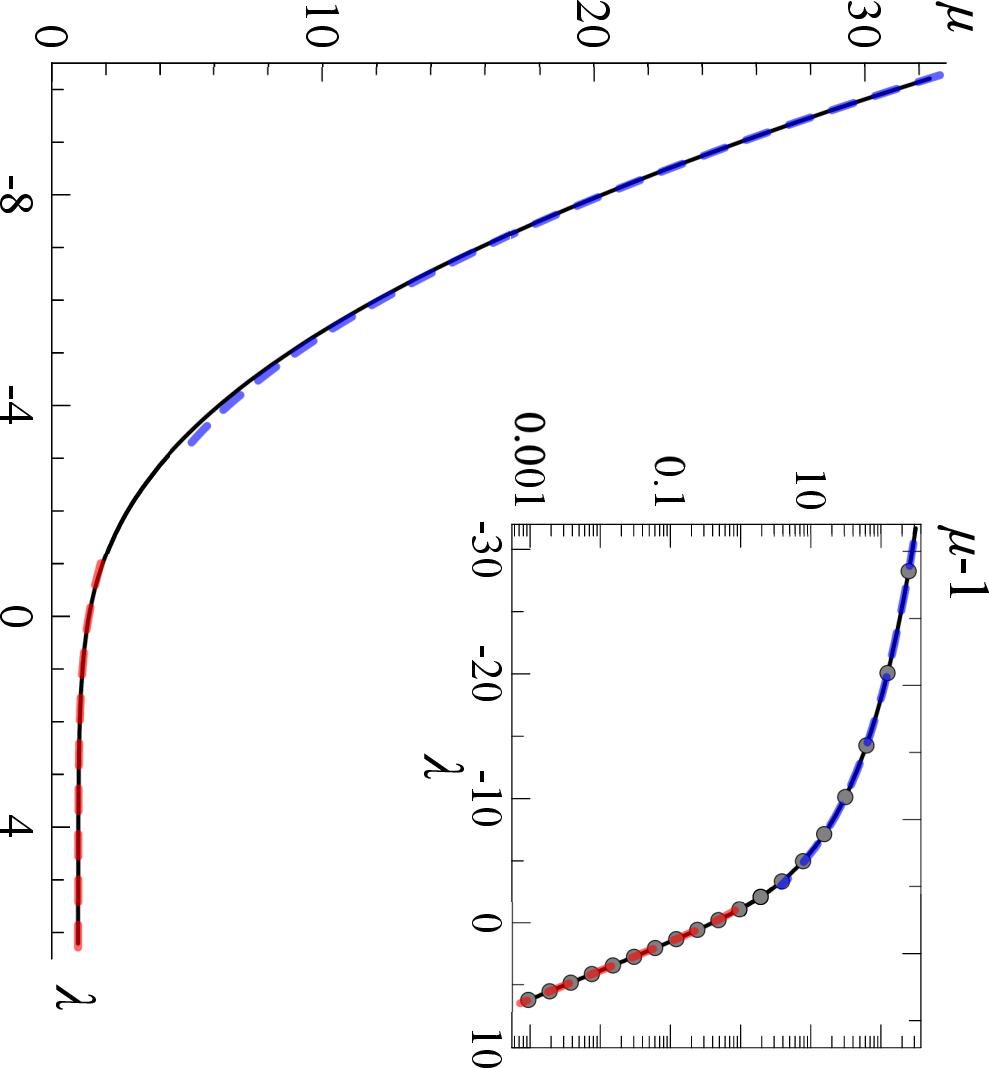}
\caption{Surface profile of the film entrained by a moving boundary. Function $\mu(\lambda)$ is shown both in linear and logarithmic (inset) scales; symbols are the results of numerical evaluation, black lines render their interpolation; color lines show the dependencies corresponding to approximations: (\ref{eq:A9}) for the `near' zone, (\ref{eq:A11}) for the `far' zone after their splicing with the numerical solution, i.e., using the obtained values of $\alpha_{-1}$ and $\alpha_1$.
\label{fig:03}}
\end{figure}

     The solution of equation (\ref{eq:17}) for the film flow in the `far' zone is presented in Appendix A.
     The evaluated there film profile (in dimensionless variables $\lambda$ and $\mu$) is shown in Figure \ref{fig:03}.
     Evidently, only the parts of the curves, which begin from around $x=0$ and go to the right, may be related to the observable data.
     In the next section the solution of Appendix A is spliced with the profile of the film in the `near' zone thus providing a consistent description of the flow over all the interval of distances.

\subsection{Film profile in the `near' zone  \label{sec:3.2}}
     In the `near' zone, that is at small values of $x$, the surface profile almost coincides with the static meniscus.
     When evaluating its shape, one may neglect viscous stresses ($\sim\eta v_0/h_0$) in comparison with the capillary pressure $\sim\sigma_0/h_0$.
     The corresponding ratio yields a criterion
\begin{equation}  \label{eq:20}
   \eta v_0/\sigma\ll1,
\end{equation}
that is independent of the gap height.

\begin{figure*}[ht]
\includegraphics[angle=90,width=0.66\textwidth]{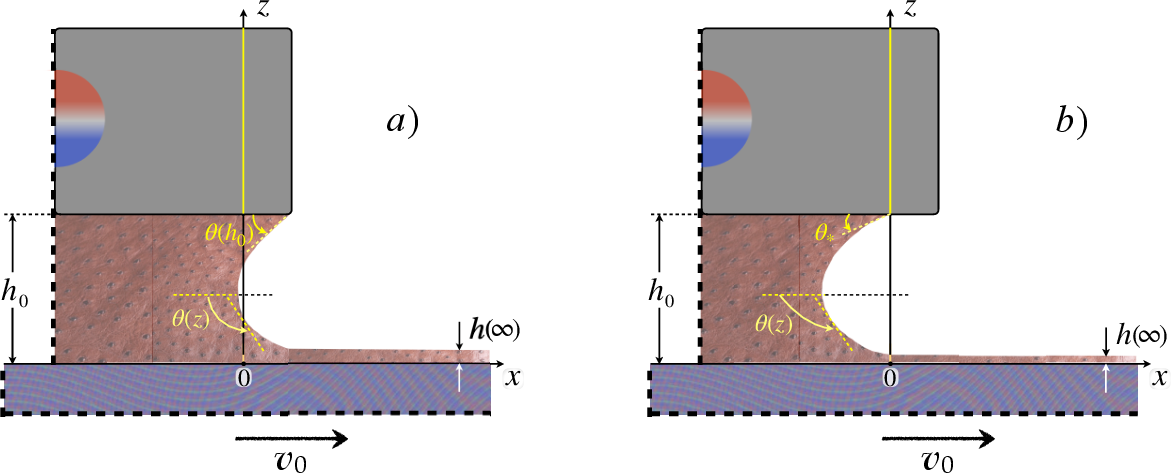}
\caption{Scheme of the fluid bridge (rear part); pane \emph{a}) -- situation of arbitrary drain; \emph{b}) -- situation of minimal drain: contact angle $\theta_0$ equals wetting angle $\theta_\ast$.
\label{fig:04}}
\end{figure*}

     It is convenient to express the curvature radius $R$ of the profile---see Figure \ref{fig:04}---in terms of surface inclination angle $\theta$ and length $l$ of the profile counted from the point where the meniscus abuts on the upper border of the gap:
\begin{equation*}
\frac1R=\frac{d\theta}{dl}=\frac{dz}{dl}\frac{d\theta}{dz}=-\sin\theta\frac{d\theta}{dz}=\frac{d\cos\theta}{dz}.
\end{equation*}
Then, setting in (\ref{eq:06}) the sub-surface pressure to the sum of capillary and magnetic pressure jumps, one finds
\begin{equation}  \label{eq:21}
   -\frac1R=-\frac{d\cos\theta}{dz}=\frac{p_0-\rho gz+\tfrac12\mu_0\chi H^2}{\sigma}+\frac{\mu_0\chi^2}{2\sigma}\frac{(H_n^{\rm(e)})^2}{(1+\chi)^2},
\end{equation}
where $p_0$ is to be evaluated by means of splicing of the film profile in the `far' zone with that of the stationary meniscus.

     Within the induction-free approximation, i.e., assuming $\chi\ll1$, one may neglect the quadratic in $\chi$ terms in comparison with the linear ones $\chi^2(H_n^{\rm(e)})^2\ll\chi H^2$ and treat the magnetic field as independent of the shape of the MF surface.
     Then equation (\ref{eq:21}) takes the form
\begin{equation}  \label{eq:22}
\frac1R=\frac{d\cos\theta}{dz}=-\frac{p_{00}}{\sigma}+\frac{\rho gz}{\sigma}-\frac{\mu_0\chi}{2\sigma}\left(x\frac{\partial H^2}{\partial x}+z\frac{\partial H^2}{\partial z} \right);
\end{equation}
here function $H^2(x,z)$ has been linearized in the vicinity of the coordinate origin, and the constant is defined as $p_{00}=p_0+\tfrac12\mu_0\chi H^2(0,0)$; the derivatives of $H^2$ are as well evaluated in the vicinity of the coordinate framework origin.

     Upon renormalizing the force parameters as
\begin{equation}  \label{eq:23}
   g_z\!=\!g\!-\!\frac{\mu_0\chi}{2\rho}\frac{\partial H^2}{\partial z}, \;   g_x\!=\!-\frac{\mu_0\chi}{2\rho}\frac{\partial H^2}{\partial x}, \;
   g^\prime\!=\!\sqrt{g_x^2+g_z^2},
\end{equation}
and tilting the coordinate framework by angle
\begin{equation}  \label{eq:24}
\theta_g={\rm arctan}(g_x/g_z),
\end{equation}
it is convenient to introduce new coordinates according to
\begin{equation*}
   x^\prime=x\cos\theta_g-z\sin\theta_g, \qquad    z^\prime=x\sin\theta_g+z\cos\theta_g.
\end{equation*}
As seen from these formulas, $g_x, g_z, \theta_g$ are now functions of the internal field strength $H$.

     In those terms, equation (\ref{eq:22}) acquires the form
\begin{equation}  \label{eq:25}
\frac{d\cos\left(\theta(z^\prime)+\theta_g\right)}{dz^\prime}=-\frac{p_{00}}{\sigma}+\frac{\rho g^\prime z^\prime}{\sigma}.
\end{equation}
     Its integration from the uppermost point of the meniscus ($z_0^\prime=h_0\cos\theta_g$, $\theta_0=\theta(h_0)$) yields
\begin{multline}  \label{eq:26}
\cos\left(\theta_0+\theta_g\right)-\cos\left(\theta(z^\prime)+\theta_g\right) \\
=\frac{p_{00}(z^\prime-z_0^\prime)}{\sigma}-\frac{\rho g^\prime(z^{\prime2}-z_0^{\prime2})}{2\sigma}.
\end{multline}

     In what follows, it is convenient to present equation (\ref{eq:26}) in a more compact form upon grouping up several terms into a single integration constant:
\begin{align}  \label{eq:27}
\cos\left(\theta(z^\prime)+\theta_g\right)=C-\frac{p_{00}}{\sigma}z^\prime+\frac{\rho g^\prime z^{\prime2}}{2\sigma}, \\
C\equiv\cos\left(\theta_0+\theta_g\right)+\frac{p_{00}}{\sigma}z_0^\prime-\frac{\rho g^\prime z_0^{\prime2}}{2\sigma}. \notag
\end{align}
     Solving this quadratic (with respect to $z^\prime$) equation, one gets
\begin{equation}  \label{eq:28}
z^\prime=\frac{p_{00}}{\rho g^\prime}\pm\sqrt{\left(\frac{p_{00}}{\rho g^\prime}\right)^2-\frac{2\sigma}{\rho g^\prime}\left[C-\cos\left(\theta(z^\prime)+\theta_g\right)\right]}.
\end{equation}

     At the bottom boundary ($z=0$) of the considered domain there must be $\theta=\pi$.
     However, solution (\ref{eq:28}) is expressed in terms of variable $z^\prime$ (tilted coordinate framework) that at $z=0$ is non-zero and is related to $x^\prime$ as
\begin{equation}  \label{eq:29}
z^\prime=x^\prime\tan\theta_g.
\end{equation}
     To evaluate $x^\prime$, we note that along the meniscus profile $dz^\prime=\tan\left(\theta(z^\prime)+\theta_g\right)\cdot dx^\prime$, so that
\begin{multline}  \label{eq:30}
dx^\prime=\frac{dz^\prime}{\tan\left(\theta(z^\prime)+\theta_g\right)} \\
=\frac{dz^\prime}{d\cos\left(\theta(z^\prime)+\theta_g\right)}\frac{d\cos\left(\theta(z^\prime)+\theta_g\right)}{\tan\left(\theta(z^\prime)+\theta_g\right)} \\
=\sigma\frac{\cos\left(\theta(z^\prime)+\theta_g\right)\cdot d\theta}{p_{00}-\rho g^\prime z^\prime},
\end{multline}
where the derivative is substituted from (\ref{eq:25}).
     Eliminating $z^\prime$ from (\ref{eq:30}) with the aid of (\ref{eq:28}), one gets
\begin{equation}  \label{eq:31}
dx^\prime=\mp\sqrt{\frac{\sigma}{2\rho g^\prime}}\frac{\cos\left(\theta+\theta_g\right)\cdot d\theta}{\sqrt{C_1+\cos\left(\theta+\theta_g\right)}}, \;\; C_1\equiv\frac{p_{00}^2}{2\sigma\rho g^\prime}-C\geq1;
\end{equation}
from now on, argument $z^\prime$ of $\theta$ is omitted.

     For the considered part of the profile $x^\prime(\theta)$ (the rear part of the fluid bridge) the `minus' sign should be chosen.
     Note that from (\ref{eq:31}), provided $dx^\prime$ is real, it follows that for arbitrary values of $\cos\left(\theta+\theta_g\right)$ the integration constant may not be below unity: $C_1\geq1$.
     The limit $C_1=1$ is also admissible, this case takes place, in particular, for a meniscus of quiescent fluid in the absence of magnetic field ($\theta_g=0$).
     In such a situation the fluid surface far from the meniscus is horizontal, i.e., at $x\rightarrow+\infty$ angle $\theta$ tends to $\pi$ from below whereas the derivative $dx/d\theta$ tends to infinity.

     The result of integration of the `minus' branch of equation (\ref{eq:31}) may be expressed in terms of elliptic integrals as
\begin{multline}  \label{eq:32}
x^\prime=C_2+\sqrt{\frac{2\sigma}{\rho g^\prime}}\left[\sqrt{C_1-1}\cdot{\rm E}\left(\cos\frac{\theta+\theta_g}{2};\frac{i\sqrt2}{\sqrt{C_1-1}}\right) \right.\\
\left. -\frac{C_1}{\sqrt{C_1-1}}\cdot{\rm F}\left(\cos\frac{\theta+\theta_g}{2};\frac{i\sqrt2}{\sqrt{C_1-1}}\right)\right],
\end{multline}
where $C_2$ is the integration constant and ${\rm E}(z;k)$ and ${\rm F}(z;k)$ are the incomplete elliptic integrals of the first and second kinds, respectively \cite{HMF_65}.

     As already mentioned, the case $C_1=1$ corresponds to a truly static ($v_0=0$) meniscus under zero magnetic field.
     However, from applicational viewpoint, interesting are the regimes where $C_1$ is substantially greater than unity.
     Given that, formula (\ref{eq:32}) may be simplified; to the first order correction with respect to the second arguments in elliptic integrals, one finds
\begin{multline}  \label{eq:33}
x^\prime\approx C_2-\sqrt{\frac{2\sigma}{\rho g^\prime}}\left[\frac{\sin(\theta+\theta_g)}{2\sqrt{C_1-1}} \right.\\
\left. -\frac{2\sin(\theta+\theta_g)+\tfrac12\sin2(\theta+\theta_g)+\theta}{8(C_1-1)^{3/2}}\right].
\end{multline}

     Taking into account that the coordinates of the uppermost point of the meniscus are $\left(x_0^\prime,z_0^\prime\right)=\left( -h_0\sin\theta_g, h_0\cos\theta_g\right)$ and using equations (\ref{eq:26}), (\ref{eq:27}) and (\ref{eq:32}), for the lowest point $\left(x^\prime,z^\prime\right)$ of the `near'-zone segment of the meniscus---where equation (\ref{eq:29}) holds---one gets the set of equations
\begin{widetext}
\begin{align}
   &\cos(\theta_0+\theta_g)+\cos\theta_g=\frac{p_{00}(z^\prime-h_0\cos\theta_g)}{\sigma}-\frac{\rho g^\prime(z^{\prime2}-h_0^2\cos^2\theta_g)}{2\sigma},
    \label{eq:34} \\
   &z^\prime=\tan\theta_g\sqrt{\frac{2\sigma}{\rho g^\prime}}\left[\sqrt{C_1-1}\cdot{\rm E}\left(\cos\frac{\theta+\theta_g}{2};\frac{i\sqrt2}{\sqrt{C_1-1}}\right)
   -\frac{C_1}{\sqrt{C_1-1}}\cdot{\rm F}\left(\cos\frac{\theta+\theta_g}{2};\frac{i\sqrt2}{\sqrt{C_1-1}}\right)\right]\bigg|_{\theta_0}^\pi-\frac{h_0\sin^2\theta_g}{\cos\theta_g},
    \label{eq:35} \\
   &C_1=\frac{p_{00}^2}{2\sigma\rho g^\prime}-\cos(\theta_0+\theta_g)-\frac{p_{00}h_0\cos\theta_g}{\sigma}+\frac{\rho g^\prime h_0^2\cos^2\theta_g}{2\sigma},
    \label{eq:36}
\end{align}
\end{widetext}
wherefrom one finds the values for $z^\prime,\,C_1,p_{00}$ at a given angle $\theta_0$.

\section{Splicing of the solutions and the fluid drain \label{sec:4}}
     In Appendix A a solution is obtained of equation (\ref{eq:17}) that describes the film flow in the `far' zone.
     Expression (\ref{eq:A10})
\begin{equation}  \label{eq:37}
\frac1{R_0}=\frac{\alpha_1^2}{2}\left(\frac{3\eta}{\sigma}\right)^{2/3}\frac{v_0^{5/3}}{j},
\end{equation}
define the curvature radius of the fluid profile at the point where it borders on the meniscus; the constant $\alpha_1=-1.1340550\ldots$ is given by (\ref{eq:A8}).
     As in the `far' zone the slope of the surface is assumed to be small ($\theta\rightarrow\pi$) but there are no restrictions on the capillary and viscous contributions, this solution might be considered appropriate for the periphery of the meniscus at the outer edge of the `near' zone where the profile slope is small as well.
     Assuming that in this region the thickness of the fluid layer is already much smaller than $h_0$, i.e., $z\rightarrow0$, one may set the curvature radii equal thus splicing the surface profiles of both zones.
     Upon doing that, from equations (\ref{eq:22}), (\ref{eq:25}) and (\ref{eq:37}) one obtains
\begin{equation}  \label{eq:38}
-p_{00}+\rho g^\prime z^\prime=\frac{1}{2}\alpha_1^2\left(3\eta\right)^{2/3}\sigma^{1/3}\frac{v_0^{5/3}}{j},
\end{equation}
that determines parameter $p_{00}$ introduced in formula (\ref{eq:22}) and thus eliminates $p_{00}$ from the set of unknown quantities in (\ref{eq:34})--(\ref{eq:36}).
     We remark that in the presence of magnetic field the pressure in the splicing point depends on the $x$-coordinate of this point.
     However, in the tilted coordinate frame it depends of the pair ($x^\prime,\,z^\prime$) since---see the left-hand part of equation (\ref{eq:38})---in that frame this pressure reads $p_{00}-\rho g^\prime z^\prime$.

     Equations (\ref{eq:34})--(\ref{eq:36}) and (\ref{eq:38}) enable one to fully evaluate the profile $x(z)$ of the MF meniscus in the `near' zone under given values of velocity $v_0$ and fluid outgo (drain) $j$.
     To do that, one should integrate the tilt angle $\theta(z)$ with respect to vertical coordinate.
     However, an accurate comparison of the theoretical predictions with experimental data on the MF meniscus is hardly probable in the near future, since precise measurement of the profile is a sophisticated task.
     Much more feasible for measurements and far more important practically characteristic of the MF bridge is the dependence of expenditure $j$ of the fluid on the velocity $v_0$ of the moving plane.
     Combining equations (\ref{eq:34})--(\ref{eq:36}) with allowance for (\ref{eq:23}), (\ref{eq:24}) and (\ref{eq:38}), one arrives at the expression
\begin{widetext}
\begin{equation} \label{eq:39}
\cos(\theta_0+\theta_g)=\cos\theta_g\cdot\left[-1+\frac{1}{2}\alpha_1^2\left(\frac{3\eta}{\sigma}\right)^{2/3}\frac{v_0^{5/3}h_0}{j} +\frac{\rho g_zh_0^2}{2\sigma}\right]-\frac{(\rho g_zh_0-p_{00})z^\prime}{\sigma}-\frac{\rho g_zz^{\prime2}}{2\sigma\cos\theta_g},
\end{equation}
or, explicitly
\begin{equation} \label{eq:40}
j= \frac{\alpha_1^2}{2} \left(\frac{3\eta}{\sigma}\right)^{2/3}\frac{v_0^{5/3}h_0}{1+\frac{\cos(\theta_0+\theta_g)}{\cos\theta_g}-\frac{\rho g_zh_0^2}{2\sigma}+\frac{(\rho g_z h_0-p_{00})z^\prime}{2\sigma\cos\theta_g} +\frac{\rho g_zz^{'2}}{2\sigma\cos^2\theta_g}}.
\end{equation}
\end{widetext}

     Let an MF with given coefficients $\eta$ and $\sigma$ be subjected to a magnetic field of certain configuration; the gap hight is $h_0$, and its lower boundary moves with a constant velocity $v_0$.
     Consider the drain of MF as a function of the boundary angle $\theta_0$.
     The minimal value of the latter is the wetting angle $\theta_\ast$ inherent to the pair MF / upper solid boundary.
     Setting $\theta_0=\theta_\ast$ in formula (\ref{eq:40}), one gets expression for the minimal drain:
\begin{widetext}
\begin{equation} \label{eq:41}
j_{\rm min}= \frac{\alpha_1^2}{2} \left(\frac{3\eta}{\sigma}\right)^{2/3}\frac{v_0^{5/3}h_0}{1+\frac{\cos(\theta_\ast+\theta_g)}{\cos\theta_g}-\frac{\rho g_zh_0^2}{2\sigma}+\frac{(\rho g_z h_0-p_{00})z^\prime}{2\sigma\cos\theta_g} +\frac{\rho g_zz^{'2}}{2\sigma\cos^2\theta_g}}.
\end{equation}
\end{widetext}
     Here $z^\prime$ and $p_{00}$ are to be evaluated from equation set (\ref{eq:34})--(\ref{eq:36}) at $\theta_0=\theta_\ast$.
     Formula (\ref{eq:41}) imposes a limit from below on the MF drain in a system with particular dimensions and given configuration of the magnetic field.

     Evidently, the situation with minimal drain is of technological interest.
     In this connection, it is useful to consider the dependence of $j_{\rm min}$ on the gap height $h_0$.
     For that, let us present function (\ref{eq:41}) in the form
\begin{equation} \label{eq:42}
j_{\rm min}=J_cF\left(\xi\right),
\end{equation}
with the notations
\begin{equation} \label{eq:43}
J_c\!=\!\tfrac12 \alpha_1^2\!\left(\frac{3\eta}{\sigma}\right)^{\!\!2/3}\!\!\!R_cv_0^{5/3}, \;\;\; \xi=\frac{h_0}{R_c}, \;\;\; R_c\!=\!\left(\frac{2\sigma}{\rho g_z}\right)^{\!\!1/2}\!\!\!;
\end{equation}
here the last coefficient stands for the capillary radius.
     The functional factor in (\ref{eq:42}) is
\begin{equation} \label{eq:44}
F(\xi)=\frac{\xi}{1+\frac{\cos(\theta_\ast+\theta_g)}{\cos\theta_g}-\xi^2+\sin\theta_g\cdot f_g(\xi,\theta_g)},
\end{equation}
where the behavior of $f_g$ is characterized by the properties: $f_g(\xi,0)\!\ne\!\infty$, $f_g(\xi,\theta_g)\!\geq\!0$, and $f_g(\xi\!\ll\!1,\theta_g)\!\sim\!\xi$.
     Therefore, function $F(\xi)$ grows monotonically with $\xi$ and is minimal for thin films ($h_0\ll R_c$):
\begin{equation*}
F(\xi)\approx\frac{\xi}{1+\frac{\cos(\theta_\ast+\theta_g)}{\cos\theta_g}}.
\end{equation*}
     In this regime the MF drain is
\begin{equation} \label{eq:45}
j_{\rm min}\big|_{h_0\ll R_c}=\tfrac12 \alpha_1^2\left(\frac{3\eta}{\sigma}\right)^{2/3}\frac{v_0^{5/3}h_0}{1+\frac{\cos(\theta_\ast+\theta_g)}{\cos\theta_g}}.
\end{equation}

     Consider a numerical estimation for the case of water in contact with ordinary glass or steel:
\begin{multline*}
   \eta\!\approx\!10^{-3}\,{\rm N}\!\cdot\!{\rm s}/{\rm m}^2, \; \sigma\!\approx\!7.3\!\cdot\!10^{-2}\,{\rm N}/{\rm m}, \; \theta_\ast\!\approx\!41^\circ, \\
   \cos\theta_\ast\!\approx\!0.75, \; \theta_g=0, \; \alpha_1^2\!\approx\!1.
\end{multline*}
     Substitution in (\ref{eq:45}) gives
\begin{equation} \label{eq:46}
j_{\rm min}\big|_{h_0\ll R_c}\!\approx\!3.4\!\cdot\!10^{-2} v_0^{5/3}h_0\;[{\rm m}^2/{\rm s}].
\end{equation}
     For the capillary radius formula (\ref{eq:43}) yields $R_c\!\approx\!0.38$\,cm.
     Let us set the contact gap to be twice smaller: $h_0\!\approx\!0.2$\,cm.
     The velocity of dragging of the boundary should be sufficiently low, let it be 5\,cm/s, i.e., $v_0^{5/3}\!\approx\!6.8\!\cdot\!10^{-3}\,({\rm m/s})^{5/3}$.
     Then from (\ref{eq:46}) one has $j_{\rm min}\big|_{h_0\ll R_c}\approx4.7\!\cdot\!10^{-7}\,{\rm m}^2/{\rm s}=4.7\!\cdot\!10^{-3}$\,cm$^2$/s.
     In the considered 2D problem the drain is the volume of fluid per unit width of the film in the direction of the $Oy$ axis taken away during unit time lapse.
     The initial volume of the fluid bridge of length $\sim5$\,cm along $Ox$ and gap height 0.2\,cm is about 1\,cm$^3$.
     Therefore, the resulting drain ranges about 0.5\,wt.\%{} per second whereas the thickness of the residual film is about
\begin{equation} \label{eq:47}
   h(+\infty)=j_{\rm min}\big|_{h_0\ll R_c}/v_0\!\approx\!10^{-3}\,{\rm cm}=10\,\mu{\rm m}.
\end{equation}

\section{ Dimensionless parametric solution  \label{sec:5}}
     Solution (\ref{eq:45}) renders the MF drain for the case of thin gap $h_0\ll R_c$ ($\xi\ll1$).
     Meanwhile, the dependence $j(\xi)$ may be obtained in parametric form for finite values of $\xi$ as well.
     Let us introduce in (\ref{eq:44}) a dimensionless constant and redefine the variable as
\begin{equation} \label{eq:48}
   P_0=p_{00}R_c/\sigma, \qquad  \zeta=z^\prime/(R_c\cos\theta_g).
\end{equation}
     Then for given values of $\theta_g$, $\theta_0$ and $C_1$, i.e., the magnetic field parameter, contact angle and integration constant, equation (\ref{eq:35}) transforms to
\begin{widetext}
\begin{equation} \label{eq:49}
   \zeta=\frac{\tan\theta_g}{\sqrt{\cos\theta_g}}\left[\sqrt{C_1-1}\cdot{\rm E}\left(\cos\frac{\theta+\theta_g}{2};\frac{i\sqrt2}{\sqrt{C_1-1}}\right)
   -\frac{C_1}{\sqrt{C_1-1}}\cdot{\rm F}\left(\cos\frac{\theta+\theta_g}{2};\frac{i\sqrt2}{\sqrt{C_1-1}}\right)\right]\bigg|_{\theta_0}^\pi-\xi\tan^2\theta_g.
\end{equation}
\end{widetext}
     Upon arranging it in a compact form
\begin{equation*}
   \zeta=\frac{\zeta_0(\theta_g,\theta_0,C_1)}{\sqrt{\cos\theta_g}}-\xi\tan^2\theta_g,
\end{equation*}
equation (\ref{eq:34}) may be presented as
\begin{equation} \label{eq:50}
   \frac{\cos(\theta_0+\theta_g)}{\cos\theta_g}+1=P_0\cdot(\zeta-\xi)-\zeta^2+\xi^2.
\end{equation}
     Expressing $\cos(\theta_0+\theta_g)$ from here, substituting it into (\ref{eq:36}) and resolving the latter for $P_0$, one gets
\begin{equation} \label{eq:51}
   P_0=2\left(\zeta\pm\sqrt{C_1/\cos\theta_g-1}\right);
\end{equation}
physically meaningful is, however, only the `minus' solution since the capillary pressure beneath the meniscus is negative.
     Formula (\ref{eq:51}) enables one to eliminate $P_0$ from equation (\ref{eq:50}) that yields
\begin{equation*}
\frac{\cos(\theta_0+\theta_g)}{\cos\theta_g}+1=(\zeta-\xi)\!\cdot\!\left[\zeta-\xi-2\left(\frac{C_1}{\cos\theta_g}-1\right)^{1/2}\right],
\end{equation*}
wherefrom
\begin{widetext}
\begin{equation*}
\frac{\xi}{\cos^2\theta_g}-\frac{\zeta_0}{\sqrt{\cos\theta_g}}=\xi-\zeta=-\sqrt{C_1/\cos\theta_g-1}\pm\sqrt{\left[C_1+\cos(\theta_0+\theta_g)\right]/\cos\theta_g};
\end{equation*}
\end{widetext}
here one has to take the `plus' branch because $\xi$ is always positive, $\zeta$ is small, and the `minus' branch is always negative.
     Finally, one finds
\begin{equation} \label{eq:52}
\xi\!=\!(\cos\theta_g)^{3/2}\!\!\left(\zeta_0\!+\!\sqrt{C_1\!+\!\cos(\theta_0\!+\!\theta_g)}\!-\!\sqrt{C_1\!-\!\cos\theta_g}\right).
\end{equation}
     Therefore, the relationship between the drain (flux)
\begin{equation} \label{eq:53}
j=\frac{J_c}{2\sqrt{C_1/\cos\theta_g-1}}
\end{equation}
and the dimensionless gap height $\xi$ is defined by the pair of equations---(\ref{eq:52}) and (\ref{eq:53})---with $C_1>1$ being the parameter.

     For the minimal drain in (\ref{eq:52}) and (\ref{eq:53}) one should set $\cos\theta_0=\cos\theta_\ast$.
     In the case of an ordinary fluid ($\cos\theta_g=0$), coefficient $\zeta_0$ turns to zero, and formula (\ref{eq:52}) simplifies to
\begin{equation} \label{eq:54}
\xi=\sqrt{C_1+\cos\theta_0}-\sqrt{C_1\!-\!1}.
\end{equation}
wherefrom $\sqrt{C_1-1}=(1+\cos\theta_0-\xi^2)/(2\xi)$, so that equation (\ref{eq:53}) reduces to
\begin{equation} \label{eq:55}
j\big|_{\theta_g=0}=J_c\frac{\xi}{\xi_0^2-\xi^2}, \qquad \xi_0\equiv\sqrt{1+\cos\theta_0}.
\end{equation}

\begin{figure}[ht]
\includegraphics[angle=90,width=0.48\textwidth]{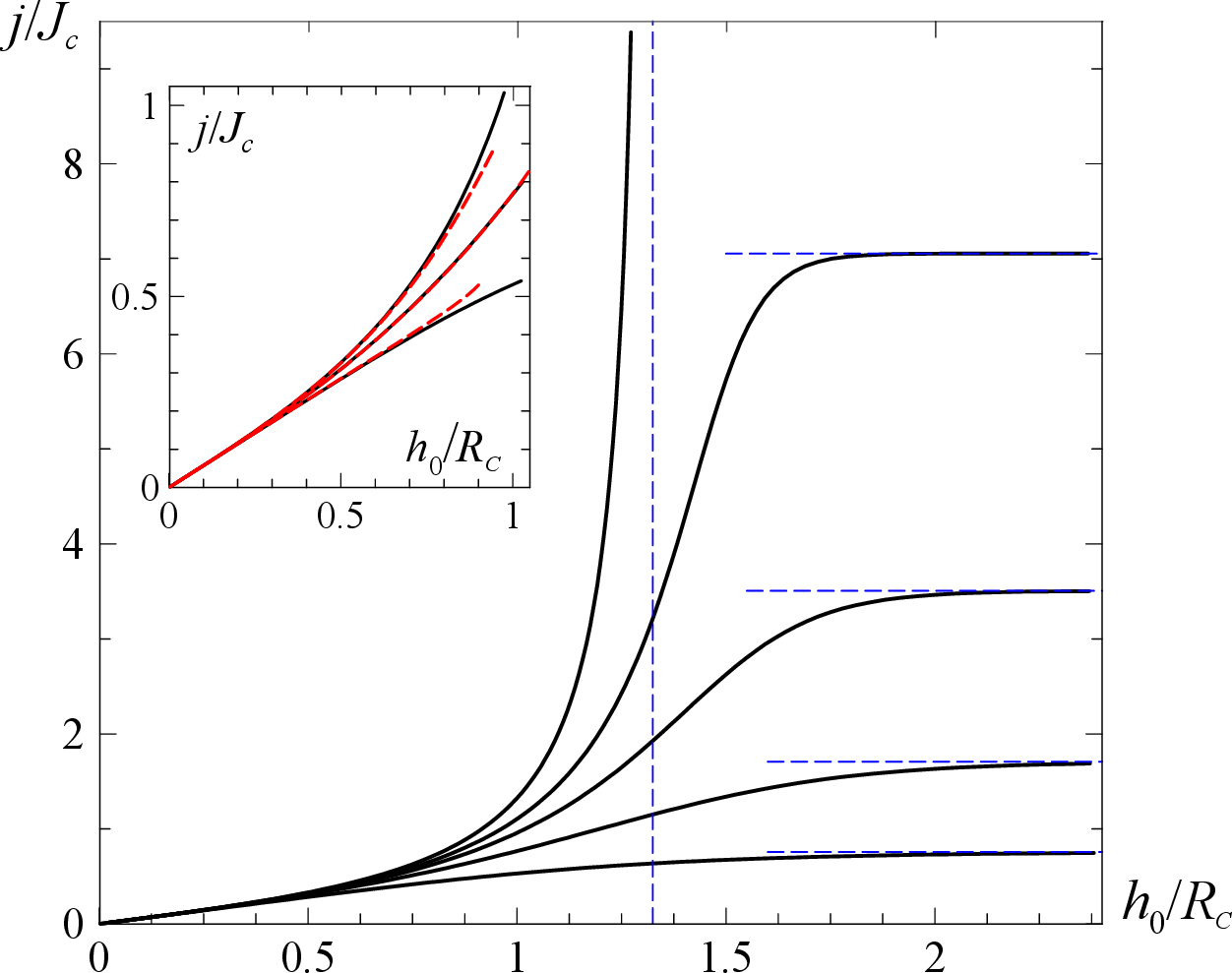}
\caption{Dimensionless drain $j/J_c$ as a function of the gap height $\xi=h_0/R_c$ for $\theta_\ast\approx41^\circ$ and $\theta_g=$0, 0.1, 0.2, 0.4 and 0.8 (from top to bottom); in the inset $\theta_g=$0.1, 0.4, 0.8.
     Solid lines show the exact solution (\ref{eq:52})--(\ref{eq:53}), dashed red lines in the inset show the results of approximation (\ref{eq:33}); vertical dashes indicate the border $\xi_\ast=\sqrt{1+\cos\theta_\ast}$, see (\ref{eq:56}); horizontal dashed lines show the maximal levels $j/J_c=0.5/\sqrt{1/\cos\theta_g-1}$ for non-zero $\theta_g$.
\label{fig:05}}
\end{figure}

     Figure \ref{fig:05} shows the behavior of function $j(h_0)$ in dimensionless form for several values of the tilt angle $\theta_g$ of mass forces, see (\ref{eq:24}).
     At a small gap height, $\xi=h_0/R_c\ll1$, the drain grows linearly for any $\theta_g$, and is well described by formula (\ref{eq:45}).
     For comparison, along with the exact parametric solution, the approximate one obtained with the aid of equation (\ref{eq:33}) is shown in the inset.

     As it is seen, in the family of curves shown in Figure \ref{fig:05} there exists a single line with a vertical asymptote that corresponds to formula (\ref{eq:55}).
     It renders the situation where magnetic field is absent ($\theta_g=0$), and the volume force is just gravity, i.e., is directed strictly vertically.
     (Evidently, under those conditions there is no difference between MF and an ordinary fluid.)

     To elucidate the cause of this divergence at $\xi\rightarrow\xi_0$, let us, first, consider the static case: $v_0=0$.
     Within the adopted approximation, in this regime the `far' zone is perfectly planar: $1/R_0=0$, that entails $p_{00}=0$ and $C_1=1$.
     Then, according to equation (\ref{eq:52}), the maximal height of the gap that is capable of retaining a fluid bridge is $\xi_0=\sqrt{1+\cos\theta_0}$, see (\ref{eq:55}) for the definition.
     Parameter $\xi_0$ grows with diminution of the boundary angle $\theta_0$ but the latter cannot be below its wetting value $\theta_\ast$.
     Therefore, the possible attainable gap height is limited by
\begin{equation} \label{eq:56}
\xi<\xi_\ast\equiv\xi_0(\theta_\ast)=\sqrt{1+\cos\theta_\ast}.
\end{equation}
     Beyond this limit, it is impossible to balance the capillary and gravity forces at the contact line, this line looses stability, and the fluid bridge falls apart; formally, the fluid drain becomes infinitely large.
     In other words, an ordinary fluid cannot be retained in a gap whose height exceeds $h_0=\sqrt{2\sigma(1+\cos\theta_\ast)/\rho g}$.

     When the bottom boundary is set to motion $v_0\ne0$, the flow of the entrained fluid induces under the meniscus surface an additional negative pressure that results in the increase of absolute value $|-1/R_0|$ of negative curvature.
     Given that, equation (\ref{eq:39}) assumes the form
\begin{equation*}
\xi^2=\left(1-J_c/j\right)+\cos\theta_0,
\end{equation*}
wherefrom it follows that a prescribed boundary angle $\theta_0$ ($\theta_\ast$ as a limit) would be attained in the gap whose height $h_0$ (or $\xi$) is smaller than that in the static case (\ref{eq:56}).

     It is worth of reminding that in the present consideration the viscous stresses are taken into account only in the `far' zone of the film.
     In the `near' zone, their effect on the meniscus shape is allowed for only implicitly via the value of the meniscus curvature in the splicing point.
     We note that, provided a full set of equations with allowance for the entire effect of viscous stresses would be applied to the problem, the increase of $j/J_c$ at $\xi\rightarrow\xi_0$ would be finite.
     However, for practically relevant ranges of the system parameters the adopted approximation holds even under substantial excess of $j$ above $J_c$.
     This implies that the diverging $j(\xi)$ dependence presented in Figure \ref{fig:05} would fairly realistically predict a drastic increase of the drain in the case of an ordinary fluid ($\theta_g=0$).

     Magnetic fluid differs from an ordinary one by its ability to be attracted to the source of a gradient magnetic field.
     A magnet positioned at the central line of the fluid bridge, as in Figures \ref{fig:01}--\ref{fig:03}, creates in the $x>0$ region a horizontal component of the volume force---parameter $g_x$, see (\ref{eq:23}) and (\ref{eq:24})---acting inside the meniscus in negative direction of $Ox$.
     The occurring attraction to the magnet is capable of diminish the velocity of MF outgo even without account for the viscous forces.
     In this case, the MF drain at $\xi>\xi_\ast$ is always finite; moreover, as it is seen in Figure \ref{fig:05}, function $j(h_0/R_c)$ quite rapidly tends to saturation.

     Evidently, there is a qualitative difference between the solution (\ref{eq:55}) with singularity and the saturating ones of equations (52)–(53), see Figure \ref{fig:05}.
     To get a notion on the character of transition between the regimes, let us consider a situation where the magnetic field contribution is small: $\theta_g\ll1$, so that $C_1=1+\varepsilon$ ($0<\varepsilon\ll1$).
     In this limit, the incomplete elliptic integrals admit logarithmic approximation \cite{HMF_65}:
\begin{equation} \label{eq:57}
\zeta_0\approx\frac{1}{\sqrt{2}}\tan\theta_g\ln\frac{1}{C_1-1}.
\end{equation}
     Then formulas (\ref{eq:23}) and (\ref{eq:24}) transform to
\begin{equation*}
\xi\approx\sqrt{1+\cos\theta_0}+\frac{\theta_g}{\sqrt{2}}\ln\frac1{e\varepsilon}, \qquad j=\frac{J_c}{\sqrt{4\varepsilon+2\theta_g^2}},
\end{equation*}
where $e$ is the base of natural logarithms.
     Combination of these relationships yields
\begin{equation} \label{eq:58}
j\approx\frac{J_c}{\sqrt{2}\theta_g}\frac{1}{\sqrt{1+\frac{2}{e\theta^2_g}e^{-\frac{\sqrt{2}}{\theta_g}(\xi-\xi_0)}}}.
\end{equation}

     Therefore, in the range $\xi>\xi_0$ where the solution with $\theta_g$ is nonexisting, does exist the solution for arbitrary small $\theta_g$ that exponentially tends to the saturation drain value $j=J_c/\sqrt{4/\cos\theta_g-4}$.
     At low values of the volume force vector tilt to vertical direction ($\theta_g\ll1$) the maximal drain is inversely proportional to it: $j\approx J_c/(\sqrt{2}\theta_g)$.

\section{Discussion   \label{sec:6}}
     Similarly to how a technological problem of coating a film with photoemulsion had inspired a purely theoretical work \cite{LaLe_ActaUSSR_42}, the problem of ultrasonic scanning with magnetofluidic contact has required to develop a dynamic description of the entrainment of magnetic fluid by a moving horizontal plane.
     As the subject of the study is a forced flow of an MF film with a free surface, this essentially distinguishes the present work from those dealing with free flows or lubrication regimes.
     In the problem under consideration, the film is dragged out from a gap of a finite height, and it turns out that the gap size is an important characteristic affecting the flow regime.
     Our analysis shows that the case of magnetic fluid is essentially different from its ordinary-fluid analogue, see Figure \ref{fig:05}.
     As seen, in the presence of just gravity (an ordinary fluid) the height of a gap capable to hold the fluid bridge is strictly limited.
     Magnetic fluid in a gap wherein a magnet produces a horizontal component of the volume force, is free from this restriction.
     Moreover, the dependence of the drain on the gap height behaves in a saturation manner.
     We surmise that this evidences the fact that, provided the gap is of sufficient size, the situation at the upper point of the meniscus ceases to affect the flow in its bottom region wherefrom the film departs.

     Let us point out the main approximations employed and present some numerical estimates aiming the latter at a reference MF of `magnetite-in-kerosene' kind with a moderate ($10\div15$\,vol.\%) content of the solid phase.
\begin{itemize}
\item[$\blacktriangleright$]
     The third term in equation (\ref{eq:16}) is omitted.
     Using expression (\ref{eq:41}) for the minimal drain, the coefficient alongside this term may be written down explicitly:
\begin{equation} \label{eq:59}
   \frac{\rho g_zj_{\rm min}^2}{(3\eta)^{2/3}\sigma^{1/3}v_0^{8/3}}\approx\frac{\alpha_1^4\rho g_z(3\eta v_0)^{2/3}h_0^2}{4\sigma^{5/3}(1+\cos\theta_\ast-h_0^2/R_c^2)^2}.
\end{equation}
     For $g_z\sim10$\,m/s$^2$ for an MF with viscosity $\eta\sim10^{-2}\,{\rm Pa}\!\cdot\!{\rm s}$ (an order higher than that of water), density $\rho\sim1.2\!\cdot\!10^3$\,kg/m$^3$ and surface tension $\sigma\sim2\!\cdot\!10^{-2}$\,N/m the value of (\ref{eq:59}) amounts to $1.5\!\cdot\!10^5\,[{\rm s}^{2/3}{\rm m}^{-8/3}]v_0^{2/3}h_0^2$.
     This turns out to be negligibly small at $h_0\sim1$\,mm and actual values of velocity (up to tens of cm/s).
\item[$\blacktriangleright$]
     The fourth term in (\ref{eq:16}) is omitted as well.
     To estimate it, let us consider its ratio to the third term for the same values of the material parameters:
\begin{multline}
 \label{eq:60}
   N\equiv\frac{\mu_0\chi j_{\rm min}^2}{6\eta v_0^3}\!\cdot\!Q\!\cdot\!\frac{(3\eta)^{2/3}\sigma^{1/3}v_0^{8/3}}{\rho g_z j_{\rm min}^2} \\
\approx\frac{\mu_0\chi v_0^{-1/3}\,({\rm m}/{\rm s})^{1/3}}{3\rho g_z}\!\cdot\!\frac{H^2}{R_H};
\end{multline}
here it is set: $\sigma/\eta\sim1$ [m/s].

     Setting $\chi\sim0.1$ and taking the distance from the center of magnet to the meniscus as $R_H\sim10^{-2}$\,m, one gets $N\sim3\!\cdot\!10^{-9}\,[{\rm m}^{7/3}\!\cdot\!{\rm s}^{-1/3}\!\cdot\!{\rm A}^{-2}]\,v_0^{-1/3}H^2$.
     Then for $v_0\sim0.1$\,m/s one finds that to condition $N\leq1$ it corresponds the field strength $H\sim1.3\!\cdot\!10^4$\,A/m or $\sim160$\,Oe, i.e., of quite a moderate magnitude.
     Apparently, due to the power-law diminution of the field with the distance from the magnet, the field effect becomes insignificant as the film moves off by several units of $R_H$.
\item[$\blacktriangleright$]
     Viscous stresses in the bulk of the steady meniscus are assumed small in comparison with the capillary pressure.
     This condition is introduced by formula (\ref{eq:20}), and for the viscosity $\eta\sim10^{-2}\,{\rm Pa}\!\cdot\!{\rm s}$ it holds for $v_0\ll2$\,m/s that is far greater than the velocity range taken for estimates in the preceding considerations.
     Condition (\ref{eq:20}) as well grants that in the adopted limit one may neglect the circular motion which is necessarily induced in a fluid bridge, one of whose boundaries undergoes motion.
\end{itemize}

     We remark that our solution is of generic kind as it does not specify neither configuration of the permanent magnet(s) nor its positioning above the gap.
     The assumption that the magnet is located in the middle section of the MF bridge is trivial since it is the MF which strives to occupy the position that is as close as possible to the source of gradient field.
     The problem statement may be easily extended to a situation where the magnet has a non-symmetrical shape or the substrate is made of a magnetic material.
     For that, one should add to the set of fluid-mechanics equations the equations of magnetostatics or low-frequency magnetodynamics if the tested item is made of a conducting material.
     Notably, as soon as a two-component acceleration $(g_x,\,g_z)$ is introduced, it becomes clear that the case of an MF film residing on an inclined plane
     is accounted for just by appropriate renormalization of those $g$-components.

\section{Conclusions  \label{sec:7}}
     A problem of a forced flow of magnetic fluid dragged by a moving boundary from a flat gap of finite height was considered.
     The fluid is retained in the gap by a gradient field of a permanent magnet forming there a fluid bridge.
     The velocity of the motion is low enough as not to induce turbulence.
     The solution was obtained as a substantial modification of the classic Landau-Levich approximation.
     It was demonstrated that in the above-mentioned regime the drain of MF from the fluid bridge obeys the classical $j\propto v_0^{5/3}$ law, the thickness of the entrained film was determined.
     
     Parameter estimations are given for the regime of minimal drain [see equation (\ref{eq:41})] complemented with (\ref{eq:34})–(\ref{eq:36}) or parametric solution (\ref{eq:49}), (\ref{eq:52}), (\ref{eq:53}), approximation (\ref{eq:45}), and Figure \ref{fig:05}].
     The applicability conditions of the developed theory are: the validity of the induction-free approximation ($\chi\ll1$), smallness conditions (\ref{eq:20}), (\ref{eq:59}), and (\ref{eq:60}).
     In Section \ref{sec:6}, these criteria are assessed and found to be met for a reference magnetic fluid of the `magnetite-in-kerosene' type.
     For such a fluid, these criteria are also translated into conditions imposed on the physically controllable parameters, viz.\ velocity $v_0$ and magnetic field strength $H$.

\section*{Acknowledgements}
This work was carried out as a part of major scientific project funded by the RMES grant (Agreement No. 075-15-2024-535 dated 23 April 2024).

\section*{Data availability statement}
The data that support the findings of this study are available from the corresponding author upon reasonable request.

\appendix
\section{Viscous flow of a fluid film in the `far' zone}
     Nonlinear equation (\ref{eq:17}) has a unique solution that is non-divergent at infinity, the only possible way is to find it numerically.
     As there are no parameters, this solution is universal for a given set of boundary conditions.
     A substantial difficulty, however, is that this equation does not yield to conventional methods of numerical integration of ordinary differential equations.
     The issue is that its solution is singular at both boundaries.
     In \cite{LaLe_ActaUSSR_42} the authors, not obtaining the solution explicitly, had calculated its certain quantitative characteristic that is necessary to obtain the drain $j$ with the accuracy of three significant digits.

     In below, besides numerical integration, the explicit (analytical) form of function $G(\mu)$ (\ref{eq:A1}) is obtained for the limiting cases.
     As well, a procedure is developed that enables one to evaluate the desired characteristics of the flow with any required accuracy.
     We begin with introduction of an auxiliary function
\begin{equation} \label{eq:A1}
G(\mu)\equiv\frac{d\lambda}{d\mu},
\end{equation}
and 
transform equation (\ref{eq:17}) to
\begin{equation} \label{eq:A2}
G^2\frac{d^2G}{d\mu^2}=-G\left(\frac{dG}{d\mu}\right)^2-\frac{\mu-1}{\mu^3}.
\end{equation}

     In (\ref{eq:A2}) the point $(\mu=1,\,G=0)$ corresponds to $({\lambda=+\infty,\,\mu=1})$ in original variables.
     However, the derivative $dG/d\mu$ is yet-to-be-determined.
     Due to that, it is impossible to commence numerical integration of equation (\ref{eq:A2}) from this point as the coefficient ahead of the higher-order derivative turns to zero.

     To circumvent this issue, we represent function $G(\mu)$ by a power series that obeys conditions $\mu=1,\,G=0$.
     A plausible variant is
\begin{equation} \label{eq:A3}
G(\mu)=\sum\nolimits_{n=1}^{+\infty}G_n(\mu-1)^n.
\end{equation}
     Substituting it in (\ref{eq:A2}) and collecting the terms with the same powers of $(\mu-1)$, one gets the relationships between coefficients $G_n$ of different indices:
\begin{multline*} \label{eq:A4}
\sum\limits_{n_1+n_2\leq n+1}\left[\left(n-n_1-n_2+1\right)\left(n-n_1-n_2+2\right)+n_1 n_2\right]
\\ \times G_{n-n_1-n_2+2}G_{n_1}G_{n_2}=\frac{(-1)^n n(n+1)}{2},
\end{multline*}
which may be presented by recurrence formulas:
\begin{equation} \label{eq:A4}
G_1=-1 \qquad {\rm for} \quad n=1,
\end{equation}
\begin{widetext}
\begin{gather}
G_n=\frac{\frac{(-1)^n n(n+1)}{2}-\sum\nolimits_{\substack{3\leq n_1+n_2\leq n+1 \\ n_1\ne n,\,n_2\ne n}}\,C_{n_1,n_2}^nG_{n-n_1-n_2+2}G_{n_1}G_{n_2}}{n^2+n+1} \quad {\rm for} \quad n\geq2,
\label{eq:A5} \\
C_{n_1,n_2}^n=(n-n_1-n_2+1)(n-n_1-n_2+2)+n_1 n_2. \notag
\end{gather}
\end{widetext}
     In particular, one gets
\begin{equation*}
G_2=\tfrac37, \;\; G_3=-\tfrac{186}{637}, \;\; G_4=\tfrac{3040}{13377}, \;\; G_5=\tfrac{48439}{256711},\;\dots\,.
\end{equation*}

     Series (\ref{eq:A3}) converges well for $(\mu-1)<0.6$ but the convergence is drastically slowed down for $(\mu-1)\approx0.8$ and completely lost for $(\mu-1)\approx1$.
     Given that, to practically calculate function $G(\mu)$, one, first, should use series (\ref{eq:A3}) to evaluate $G$ and $dG/d\mu$ with demanded accuracy in some internal point of the interval.
     For instance, at $\mu=1.4$
\begin{equation} \label{eq:A6}
\begin{array}{r}
\displaystyle
G(1.4)
=-0.73810034099866410\ldots , \\[5pt]
\displaystyle
\frac{dG}{d\mu}(1.4)
=-0.57301393905742418\ldots .
\end{array}
\end{equation}
     The obtained values are to be used as the initial conditions for numerical integration of equation (\ref{eq:A2}) towards increasing $\mu$'s.
     This enables one to find $G(\mu)$ with arbitrary accuracy beyond the convergence interval of series (\ref{eq:A3}) and up to sufficiently large, although finite, values of $\mu$.

     For asymptotically high values of $\mu$, equation (\ref{eq:A2}) admits expansion in the form
\begin{equation} \label{eq:A7}
G(\mu)=\alpha_1\sqrt{\mu}+\frac{\alpha_{-1}}{\mu^{1/2}}+\frac{\alpha_{-2}}{\mu}+\frac{\alpha_{-3}}{\mu^{3/2}}+\dots\,,
\end{equation}
that starts from the element $\propto\sqrt{\mu}$.
     Note that the element with $\mu^0$ is absent; in other words, $\alpha_0=0$.
     Coefficients $\alpha_1$ and $\alpha_{-1}$ here are yet undetermined but as soon as they are evaluated, all those with higher indices will be determined via the recurrence relations dictated by the substitution of (\ref{eq:A7}) into (\ref{eq:A2}):
\begin{multline*}
\alpha_{-2}=-\frac{4}{3\alpha_{1}^2}, \quad \alpha_{-3}=-\frac{\alpha_{-1}^2}{2\alpha_{1}}, \\ \alpha_{-4}=\frac{4\alpha_1+24\alpha_{-1}}{15\alpha_{1}^3}, \quad \alpha_{-5}=\frac{9\alpha_1^3\alpha_{-1}^3-20}{18\alpha_{1}^5}, \ldots .
\end{multline*}

\begin{figure}[ht]
\includegraphics[angle=90,width=0.4\textwidth]{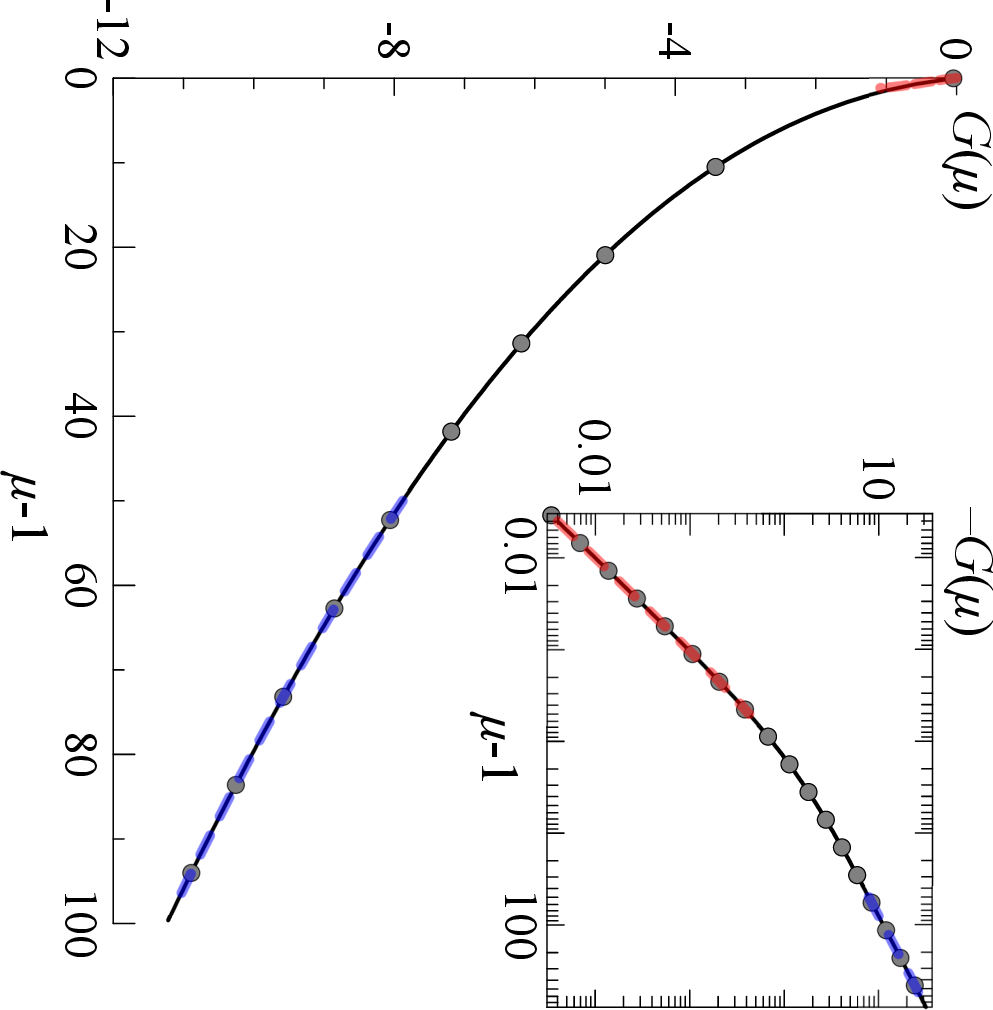}
\caption{Function $G(\mu)$, see (\ref{eq:A1}), characterizing the profile of the film entrained by a moving boundary.
     The function is shown both in linear and logarithmic (inset) scales; symbols are the results of numerical evaluation, black lines render their interpolation; color lines show the dependences corresponding to approximations: for large (\ref{eq:A7}) and small (\ref{eq:A3}) values of $(\mu-1)$ after their splicing with the numerical solution, that is using the obtained values of $\alpha_{-1}$ and $\alpha_1$.
\label{fig:06}}
\end{figure}

     By way of splicing the functions $G$ and $dG/d\mu$, given by asymptotic expansion (\ref{eq:A7}), with the numerical solution in the intermediate range, see Figure \ref{fig:06}, one can evaluate $\alpha_1$ and $\alpha_{-1}$ with any required accuracy.
     In particular,
\begin{equation} \label{eq:A8}
   \alpha_1=-1.1340550... , \qquad \alpha_{-1}=1.6442... .
\end{equation}

     As a result, function $G(\mu)$ gets defined along the half-line $\mu (1,\in +\infty)$: for small $(\mu-1)$ it is rendered by series (\ref{eq:A3}), for finite $\mu$'s by the numeric solution of (\ref{eq:A2}) with initial conditions (\ref{eq:A6}), and for high $\mu$'s by asymptotic series (\ref{eq:A7}) with coefficients (\ref{eq:A8}).
     Function $\mu(\lambda)$ is retrieved from $G(\mu)$ by integration of expression (\ref{eq:A1}); Figure \ref{fig:03} illustrates its behavior.
     For practical purposes, however, it suffices to know the dependence $\mu(\lambda)$ at small $\lambda$ where $\mu$ is large, so that the film profile is defined by the leading term of series (\ref{eq:A7}).
     Inserting the truncated asymptotic series
\begin{equation*}
   \frac{d\mu}{d\lambda}=\alpha_1\sqrt{\mu}+\frac{\alpha_{-1}}{\sqrt{\mu}}+O\left(\frac{1}{\mu}\right)
\end{equation*}
into (\ref{eq:A1})
for the region where this (`far'-zone) profile matches that of the capillary meniscus, one has
\begin{equation} \label{eq:A9}
   \mu(\lambda)=\frac{\alpha_1}{4} \left(\lambda-\lambda_0\right)^2+2\frac{\alpha_{-1}}{|\alpha_1|}+O\left(\frac{1}{\sqrt{\mu}}\right),
\end{equation}
where $\lambda_0$ is the integration constant that determines the shift of the profile along $Ox$.

     Double differentiation of (\ref{eq:A9}) renders the curvature radius $R_0$ of the film close to the capillary meniscus.
     Using the scaling (\ref{eq:15}), one gets
\begin{equation} \label{eq:A10}
 \frac{1}{R_0}\!=\!\frac{d^2h}{dx^2}\!=\!\left(\frac{3\eta}{\sigma}\right)^{\!\!\!\!2/3}\frac{v_0^{5/3}}{j}\frac{d^2\mu}{d\lambda^2}\Big|_{\mu\rightarrow\infty}\!=\!\frac{\alpha_1^2}{2}\left(\frac{3\eta}{\sigma}\right)^{\!\!\!\!2/3}\frac{v_0^{5/3}}{j}.
\end{equation}

     To complete the profile evaluation, we derive an asymptotic solution for the `far' zone where $(\mu-1)\ll1$.
     Retaining only the leading terms of series (\ref{eq:A3}), one arrives at the equation
\begin{equation*}
   \frac{d\mu}{d\lambda}=-(1-\mu)+\frac37(1-\mu)^2+O\left[(1-\mu)^3\right],
\end{equation*}
and after integration
\begin{equation} \label{eq:A11}
   \mu=1+\frac{1}{e^{(\lambda-\lambda_\infty)}+3/7}+O\left[e^{-3(\lambda-\lambda_\infty})\right],
\end{equation}
where $\lambda_\infty$ is the integration constant that determines the shift of the profile in the horizontal direction.
     The last term in (\ref{eq:A11}) is the correction of the third order of magnitude: $O\left[(1-\mu)^3\right]\sim O\big[e^{-3(\lambda-\lambda_\infty)}\big]$.
     To the leading order, $\mu$ exponentially approaches unity whereas expression (\ref{eq:A11}) provides the first correction to that dependence.
     Splicing of the solutions for `near' (\ref{eq:A9}) and `far' (\ref{eq:A11}) zones by means of the numerical solution, defines the relationship between the integration constants which define the horizontal shifts: $\lambda_\infty-\lambda_0\approx-0.14$.

Technical implementation of this procedure of computation of $\alpha_{1}$ and $\alpha_{-1}$ requires elevated machine accuracy of calculations, because of a slow convergence of expansions (\ref{eq:A3}) and (\ref{eq:A7}) for moderate values of $(\mu-1)$ and $1/\mu$, respectively, and the instability of the numerical solution of equation~(\ref{eq:A2}) for intermediate values of $\mu$. The latter instability is the inherent instability of the initial value problem, but not the numerical instability of the integration procedure. Specifically, for computation of constants (\ref{eq:A8}) with the provided number of trusted digits we employed the `Maple' package for analytical calculations with $N_d=32$ decimal digits (environmental variable `{\it Digits}'=32). We used $N_{\mu-1}=82$ elements in series (\ref{eq:A3}) to compute the initial values (\ref{eq:A6}); the numerical solution of (\ref{eq:A2}) with initial values (\ref{eq:A6}) was computed for the interval $\mu\in[1.4;401]$; the values of $G$ and $dG/d\mu$ at $\mu=401$ were equalized to the ones given by series (\ref{eq:A7}) with $N_{1/\mu}=21$ elements; the numerical solving of the last equality for $(\alpha_1,\alpha_{-1})$ yielded (\ref{eq:A8}). The provided digits in (\ref{eq:A8}) were validated to be insensitive to the further increase of accuracy and variation of the procedure parameters.

In~\cite{LaLe_ActaUSSR_42}, the first three digits of $\alpha_1$ were estimated by means of an {\it ad hoc} method, for which the enhancement of accuracy was practically unfeasible. Meanwhile, nonlinear problem (\ref{eq:17}) for natural boundary conditions has a unique solution, and these problem and solution might be of universal validity not only for the physics of film flows but also from the view point of mathematical physics. Here we provided the solution to the problem with enhanced accuracy and the procedure which allows one to conduct computations with any required accuracy.

\section*{References}
\nocite{*}
\bibliography{bib_film_flow_mag}
\end{document}